\RequirePackage{snapshot}

\documentclass[english]{revcoles}

\usepackage[tight]{subfigure}
\usepackage{siunitx}
\usepackage{booktabs}
\usepackage{hyperref}
\usepackage{rotating}
\usepackage{microtype}
\usepackage{harvard}

\begin{document}

\title[maintitle = Visualization of Skewed Data,
       secondtitle = A Tool in \texttt R,
       shorttitle = Visualization of Skewed Data
]


\begin{authors}
\author[firstname = Raydonal,
        surname = Ospina,
        numberinstitution = 1,
        affiliation = Universidade Federal de Pernambuco,
        email = rayospina@gmail.com]
\author[firstname = Antonio Marcos,
        surname = Larangeiras,
        numberinstitution = 2,
        affiliation = MSc Candidate,
        email = amlarangeiras@gmail.com]
\author[firstname = Alejandro C.,
        surname = Frery,
        numberinstitution = 2,
        affiliation = Universidade Federal de Alagoas,
        email = acfrery@gmail.com]
\end{authors}
\begin{institutions}
     \institute[
                division = Departamento de Estatística,
                institution = Universidade Federal de Pernambuco,
                city = Recife,
                country = Brazil]
     \institute[                
     			division = Laboratório de Computação Científica e Análise Numérica,
                institution = Universidade Federal de Alagoas,
                city = Maceió,
                country = Brazil]
\end{institutions}

\begin{mainabstract}
In this work we present a visualization tool specifically tailored to deal with skewed data.
The technique is based upon the use of two types of notched boxplots (the usual one, and one which is tuned for the skewness of the data), the violin plot, the histogram and a nonparametric estimate of the density.
The data is assumed to lie on the same line, so the plots are compatible.
We show that a good deal of information can be extracted from the inspection of this tool; in particular, we apply the technique to analyze data from synthetic aperture radar images.
We provide the implementation in \texttt R.
\end{mainabstract}
\section{Introduction}

\possessivecite{TukeyEDA} work set the basis for the Exploratory Data Analysis, which is the art of seeking for relevant information within the data with the least possible distributional assumptions about the underlying process.
Such quest is frequently based upon graphical representations.

Among the schematic plots which survived or emerged since the advent of powerful personal computers, one should mention, for univariate data, 
the scatterplot, 
the histogram \citeaffixed{Pearson1895Histogram}{defined and discussed in}, 
the boxplot~\cite{AdjustedBoxplotSkewed},
the beanplot~\cite{Beanplot}, 
the shifting boxplot~\cite{Marmolejo}, 
the violin plot~\cite{ViolinPlot}, 
and their many variations.

The histogram and the boxplot are the most used plots which convey information about the shape of the underlying distribution.
They work in the same fashion: they extract and display key quantifiers from the data.
These quantifiers can be tuned for specific situations as, for instance, the choice of the bins in the histograms (the Feedman-Diaconis, Sturges, and Scott options in the \verb+hist+ function available in \texttt R).

A key point when using more than a single graphical presentation of the same data set is to clearly convey the same or complementary information.
A common mistake is simply showing side-by-side several summaries, but the precision and extent of enhancement such juxtaposition provides is arguable.
Since no single plot is able to provide all the relevant information in every conceivable case, a possible solution for this problem consists of presenting the plots with clear visual clues of their same origin: the data set.

Visualization techniques are often used to drive important decisions.
If the information conveyed by the graphical summaries is hindered, decisions may be biased or completely wrong.
For instance, \citeasnoun{AutomatedNonGaussianClusteringPolSAR} present a segmentation procedure for polarimetric synthetic aperture radar (PolSAR) imagery which, albeit automatic, exhibits the quality of the product at each iteration in the form of histograms overlapped with fitted densities; the closer the fit, the better the result.
When the data is overly asymmetric, the automatic presentation is hard to grasp as the abscissas span a huge interval.

In this article we present a tool for the visual display of skewed data developed and freely available in \texttt R.
The tool is based on the integration and coordination of several graphical representations, some of them tailored to this kind of data.
We test the tool on PolSAR data, which exhibit intense asymmetry.

Section~\ref{Sec:Summaries} presents the graphical summaries that will be integrated in our visualization tool.
Section~\ref{Sec:Results} presents the data: different types of land cover as retrieved by Synthetic Aperture Radar -- SAR sensors.
This kind of data is prone to presenting extreme deviations from the Gaussian hypothesis, as they are heavily skewed.
Finally, section~\ref{Sec:Conclu} concludes the paper with further suggestions.
The Appendix provides details about the implementation and instructions to obtain the code and the data.

\section{The summaries and their coordination}\label{Sec:Summaries}

In the following we define and comment advantages and disadvantages of some commonly used graphical summaries of data.
The use of these summaries is illustrated with the same data set: a sample from a PolSAR image of the Niigata area; see Fig.~\ref{Niigata}.
As presented in Table~\ref{tab:Summary}, the VV polarization is the one with strongest asymmetry, so the data henceforth presented come from this band.
More information about this and other images is given in Section~\ref{Sec:Results}.

\subsection{Histograms and kernels}

In Data Exploratory Analysis the probability density estimation is one of the main tools to extract data information about the distribution of the underlying population.
The estimated density may help revealing patterns and features representative of the targeted object for data modeling, analysis and decision management. 

Generally speaking, the problem of density estimation can be defined as the estimating processes of the unknown distribution by means of the also unknown density function $f$ on the set of attributes of data $\mathcal{X}$ 
using the information of the observations of the random sample of size $n$, namely $\mathbb{X}=\{ X_1, \ldots, X_n \} \subset \mathcal{X}$ draw from the target function $f$. 

The Histogram is a basic form of nonparametric density estimator where the region covered by $\mathcal{X}$ is usually divided into equal-sized bins whose height is proportional to the count of hits within that bin. 
This estimator is depends on the choice of bin width $h$ and the starting points of the bin too. 
These two values will determine how the data will be grouped i.e. to which bin the data will belong to. 
The number of bins $k$ should be related to the bin width $h$ as, for instance, $ k = {\rm range}(\mathbb{X})/h$. 
Different rules to choose $h$ are available.  
For example, \citeasnoun{Scott79} proposed to $h= 3.5 \widehat\sigma/n^3$ with $\widehat\sigma$ the sample standard deviation, while \citeasnoun{Sturges26} proposed $k = 1 + \log_2(n)$. 
\citeasnoun{Freedman81} proposed to use $h=2 IQR(\mathbb{X})/n^3$  where $IQR(\cdot)$ stands for the interquartile range of the data set. These methods usually affect strongly the histogram by the start end points of the bins and their width. Additionally,  the histogram present the inconvenient feature of non-smoothness \cite{Silverman86}.

Thus, \citeasnoun{rosenblatt1956} and \citeasnoun{parzen1962} developed the kernel density estimator that is smoth, controls bin boundary effects and (under very mild conditions) also converges to the true density, but faster than the histogram. 
A ``kernel'' is any smooth function (generally, a symmetric probability density) that depends on the  bandwith parameter $h$ which controls both the spread and the orientation. 
Just like in histograms, $h$ determines the smoothness of the estimation. 
In practice,  the choice of kernel is less important than tat of $h$. 
For example, a small value of $h$ will lead to under-smoothing and masking important features of the data, such as skewness and multimodality. 
On the other hand, rough curves produced by larger values of $h$ yield smoother estimates but might dodge significant peaks or other important structure estimate \cite{Silverman86}. 

\begin{figure}[htb]
  \centering
  \includegraphics[scale=0.55]{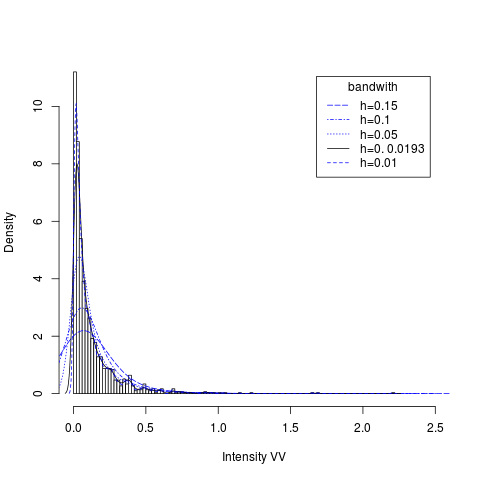}
  \caption{Histogram with kernel density estimation for the intensity VV of Niigata data set. The bandwith $h$ was chosen for different values.}\label{Fig1}
\end{figure}

Figure~\ref{Fig1} shows the histogram along with several density estimates using different bandwidths for the intensity VV of Niigata data set. 
We see how sensitive the estimate of $\widehat f$ in relation to $h$. 
Note that the graphs reveal the strong data asymmetry. 
In particular, due to the combined effect of asymmetry and large data spread, the bulk of the information is confined to a small region, approximately in the interval $[0,.7]$, while there is little to visualize in the remaining area of the plot which spans in $[.7,2.5]$.

\texttt R produces high-quality and fully customizable histograms with the \verb+hist+ function.

\subsection{Boxplots and variants} 

\citeasnoun{TukeyEDA} introduced the boxplot to analyze univariate data sets by graphically displaying important core statistics of the data.
The  plot extracts a few descriptive parameters and shows information about location, spread, skewness as well as the tails of the data. 
A boxplot shows the data distribution in terms of its quartiles, labelled $Q_1, Q_2, Q_3$ (the first, second and third quartiles).  

Define the interquartile range as $IQR=1.5(Q_3 - Q_1)$ the boxplot is comprised of the following elements:
\begin{enumerate}
\item a box, with horizontal lines at $Q_1,$ $Q_2$ (the median) and $Q_3$;  
\item vertical lines at ${W}_{L} =  Q_1 - IQR$ and ${W}_{U} = Q_1 + IQR$ (the ``whiskers'', omitted inside the box); and
\item individual observations: all observations outside the $({W}_{L}, {W}_{U})$ range (outliers), plus two observations on either en which just fall inside this range.
\end{enumerate}

A variation of the boxplot is the notched boxplot \cite{McGill} which is useful for determining whether two samples were drawn from the same population in terms of their median values.  
The notch displays a confidence interval around the median based on the Gaussian hypothesis: $Q_2 \pm 1.57\cdot IQR/\sqrt{n}$.  
According to \citeasnoun{chambers1983gmd}, although not a formal test, if two boxes notches do not overlap there is ``strong evidence'' (95\% confidence) that their medians differ. 

{\tt R}'s default graphical tools include the \verb+boxplot+ function which has the option \verb|notch=TRUE| to add a notch to the box.

In many situations, mainly when with skewed data, the boxplot may erroneously identify values which exceed the whiskers as outliers.
To correct this distortion, \citeasnoun{AdjustedBoxplotSkewed} proposed an adjusted boxplot for skewed distributions. 
The main idea is the inclusion of the {\it  medcouple} introduced by \citeasnoun{Brys} as a robust measure of skewness in the determination of the whiskers. 
Also, the adjusted boxplot can be useful as a fast tool for automatic outlier detection, without making any assumption about the distribution of the data. 
The function \verb|adjbox| of the {\tt R} package {\tt robustbase} can be used to produce an adjusted boxplot.

\begin{figure}[hbt]
\begin{center}
\subfigure[Classical boxplot\label{fig:ClassicalBoxplot}]{\includegraphics[ width=.32\linewidth]{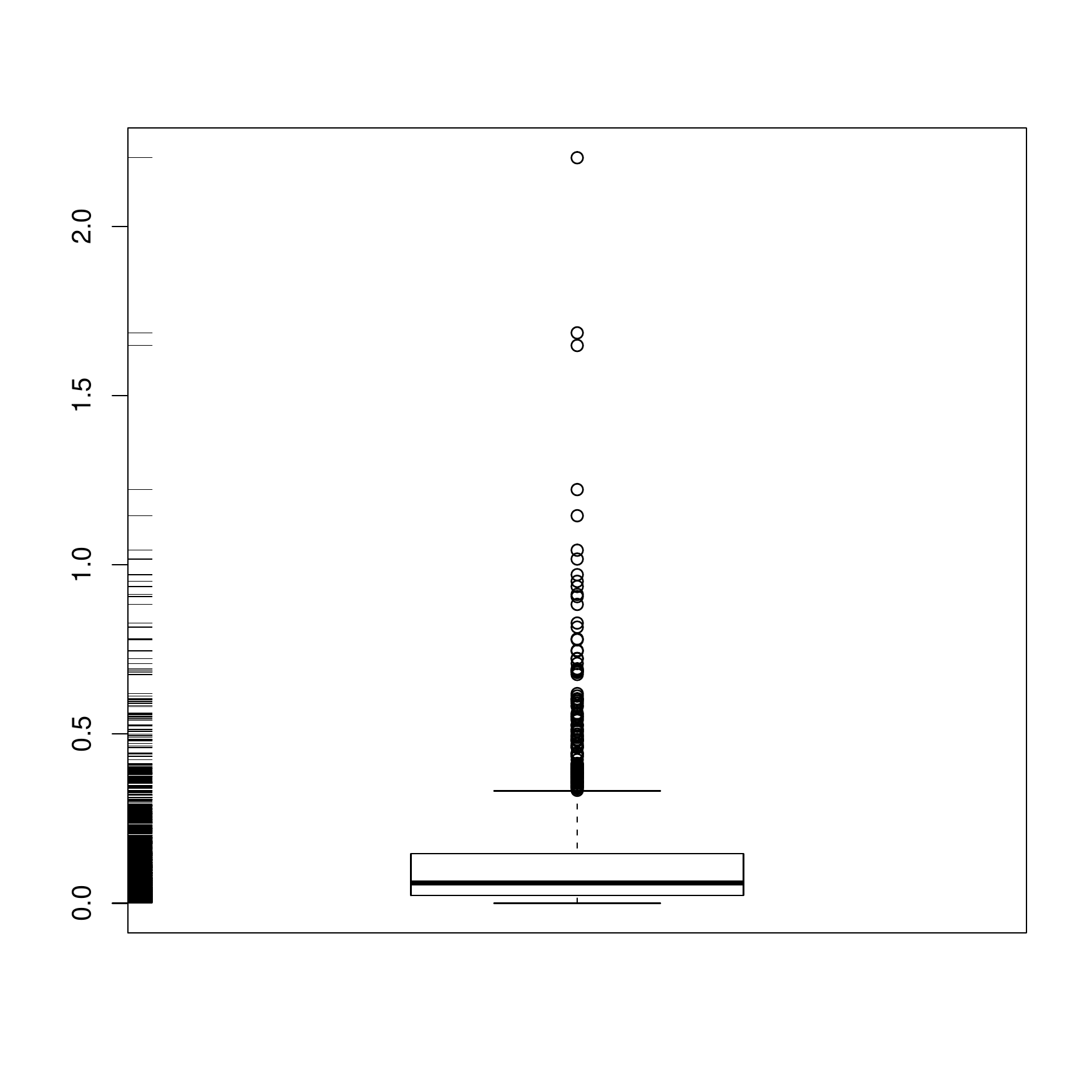}}
\subfigure[Boxplot with notches\label{fig:NotchedBoxplot}]{\includegraphics[width=.32\linewidth]{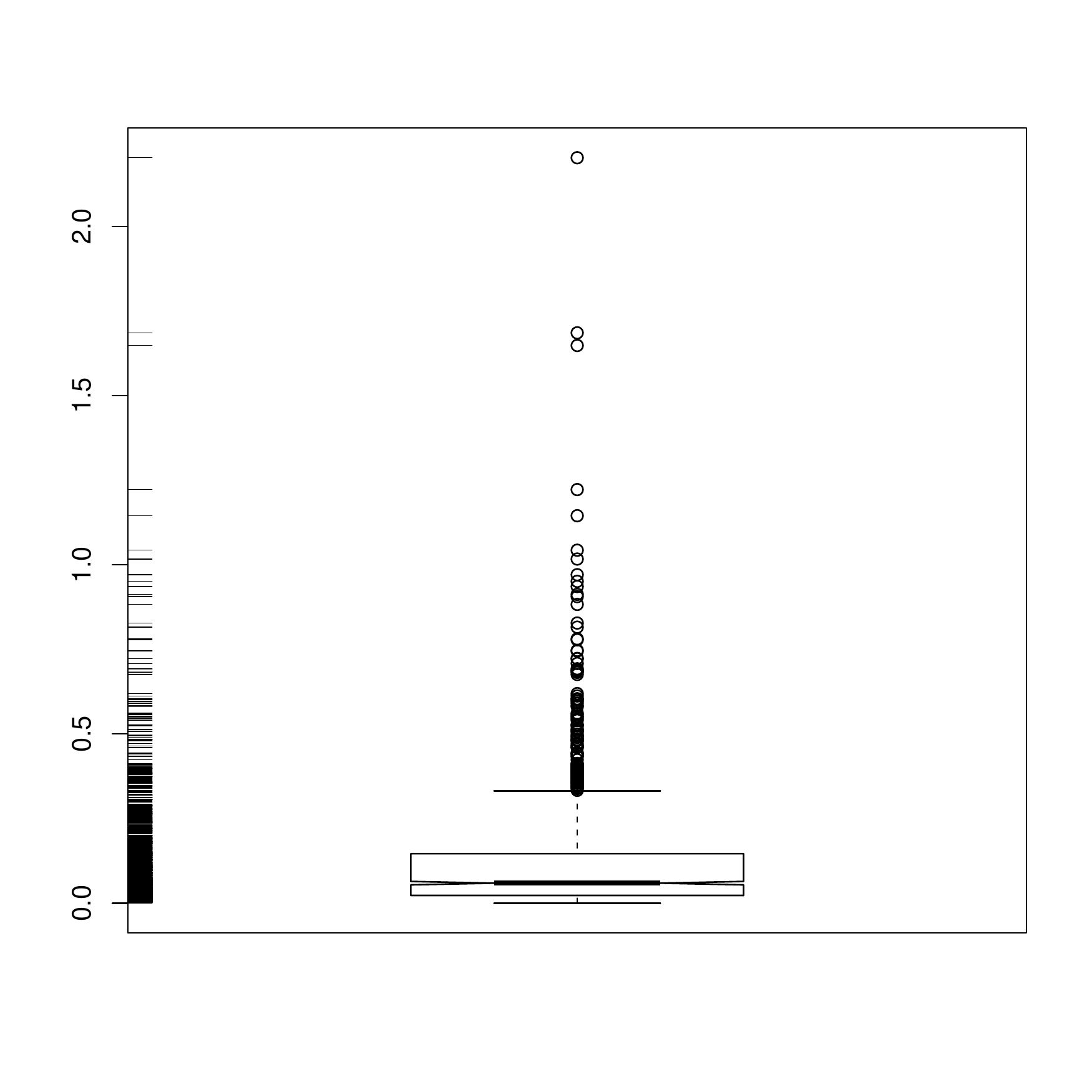}}
\subfigure[Adjusted boxplot\label{fig:AdjustedBoxplot}]{\includegraphics[width=.32\linewidth]{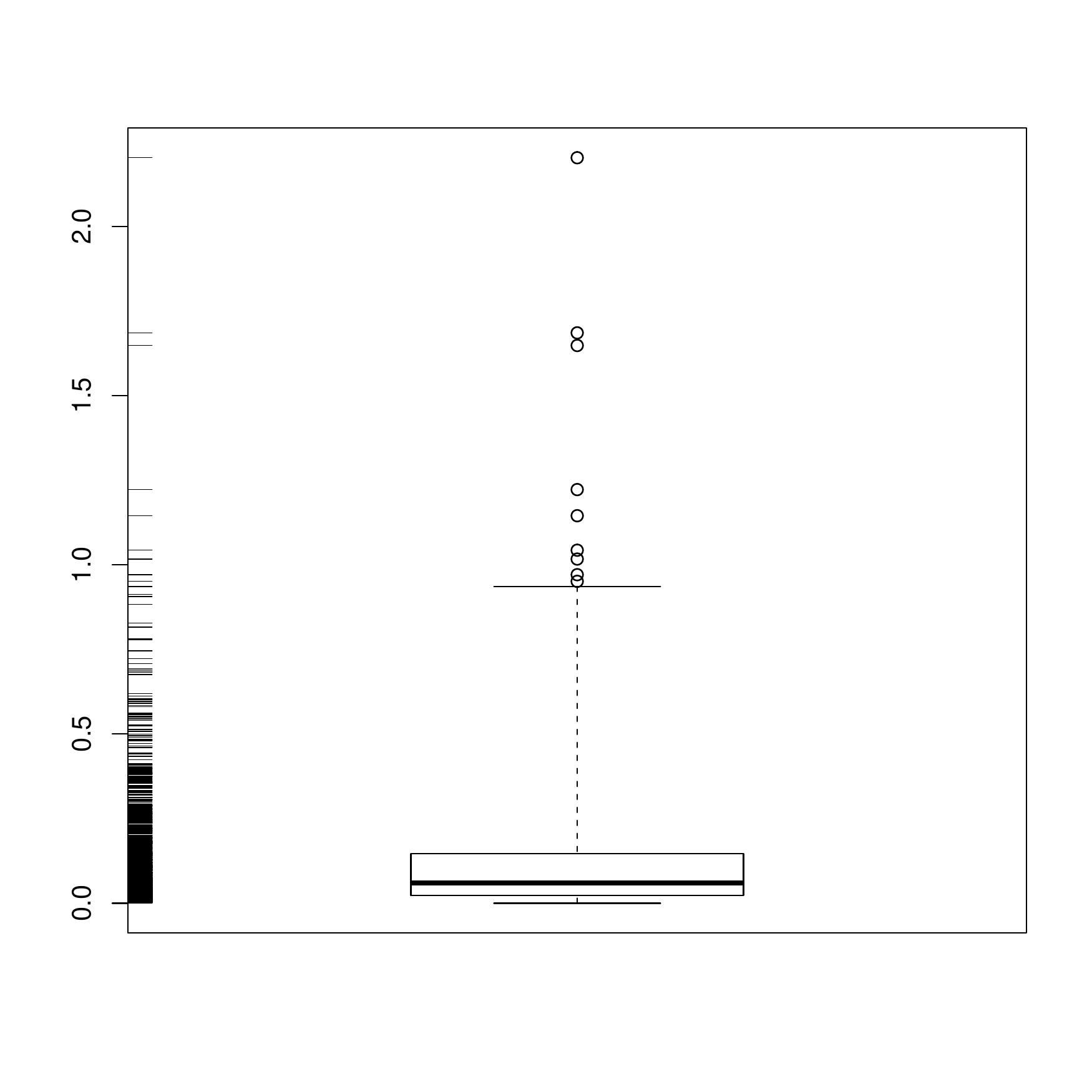}}
\caption{Boxplots for the intenssity VV of the Niigata data set.}\label{Fig2}
\end{center}
\end{figure}

Figure~\ref{Fig2} illustrates these three types of boxplot with the Niigata VV data set.
Notice that the classical and notched boxplots look alike; cf. figures~\ref{fig:ClassicalBoxplot} and~\ref{fig:NotchedBoxplot}, resp.
This is typical of large samples, $n=4446$ in this case, for which the width of the notch becomes negligible.

The difference between both classical and notched boxplots and the adjusted boxplot shown in Fig.~\ref{fig:AdjustedBoxplot} is noticeable.
While the two former identify numerous observations as outliers, the latter only considers a few as surprising data.
This last graphical representation is more adequate and may lead to better informed decisions than the former.

\citeasnoun{Warren} proposed the box-percentile plot as a variant of the boxplot which allows the sides of the plot to convey more information, presenting details about the distribution of the data. 
The width of the box is not fixed, but proportional the the number of data.
In this way, the box-percentile plot summarizes more than the histogram, but shows more details than the boxplot.
The \texttt{HMisc} package in \texttt R computes and displays this graphical summary with the function \verb|bpplot()|.

Generally speaking, boxplots are useful for small or moderate-sized data sets. 
The classical boxplot can be expected to label an increasing number of observations as outliers as the sample size grows. 
To improve the boxplot in this direction, \citeasnoun{Letter} proposed the letter-value boxplot by displaying more detailed information over the tails of distribution using letter values, but only out to the depths where the letter values are reliable estimates.
In this form  the ``outliers'' can be defined as a function of the most extreme letter value shown. 
The function \verb|LVboxplot|, implemented in {\tt R} in the appendix of \citeasnoun{Letter}, can be used to produce letter-valued boxplots. 

Recently, \citeasnoun{Marmolejo} provided a comprehensive literature review on boxplot graphs, and proposed an important variant, the Shifting boxplot. 
This graphical summary incorporates the mean as basilar information instead of the median. 
The methodology supports conducting parametric tests.  
The code that produces is can be obtained directly from the authors. 

\begin{figure}[hbt]
\begin{center}
\subfigure[Box-percentile plot]{\includegraphics[ width=.32\linewidth]{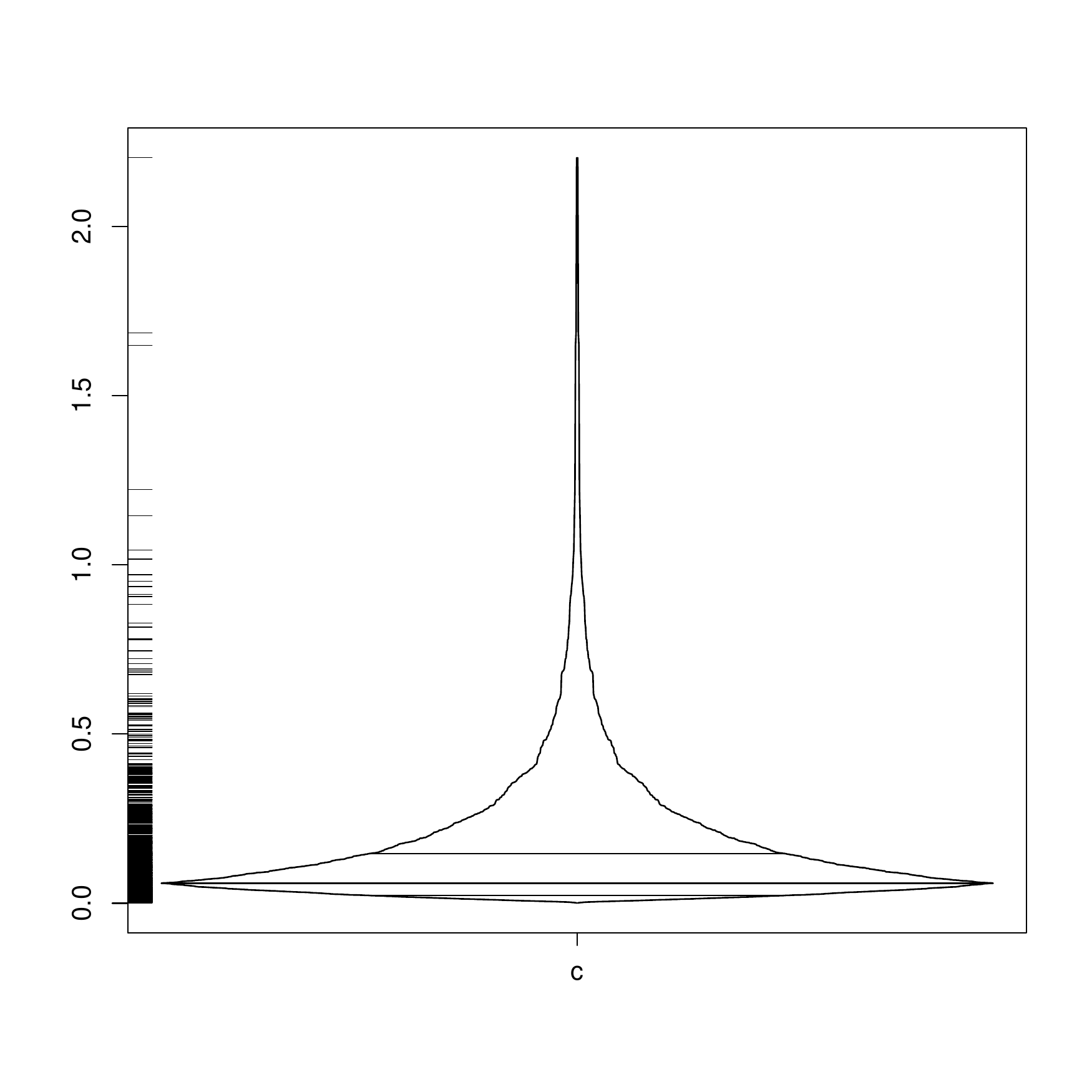}}
\subfigure[Letter-value boxplot]{\includegraphics[width=.32\linewidth]{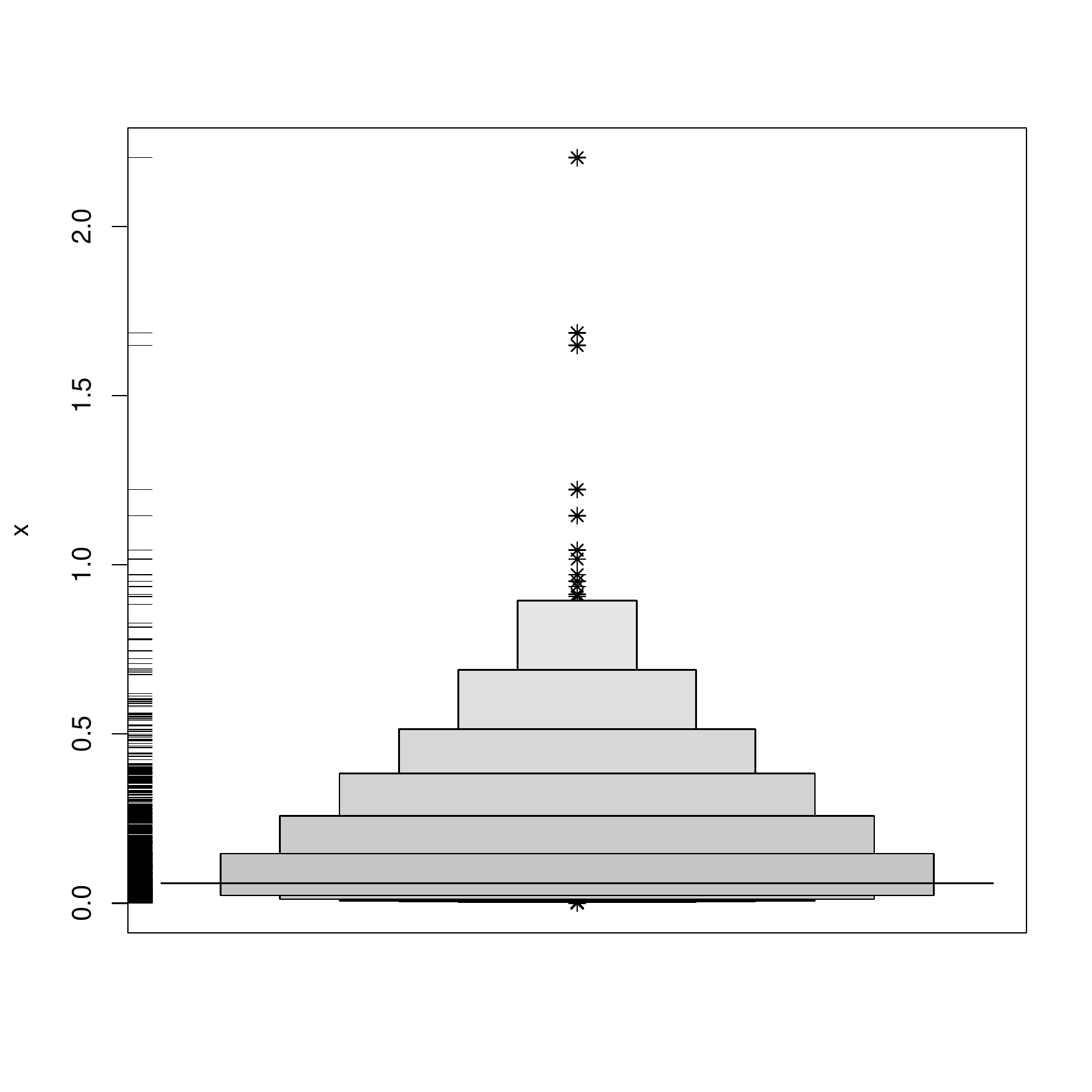}}
\subfigure[Shifting boxplot]{\includegraphics[width=.32\linewidth]{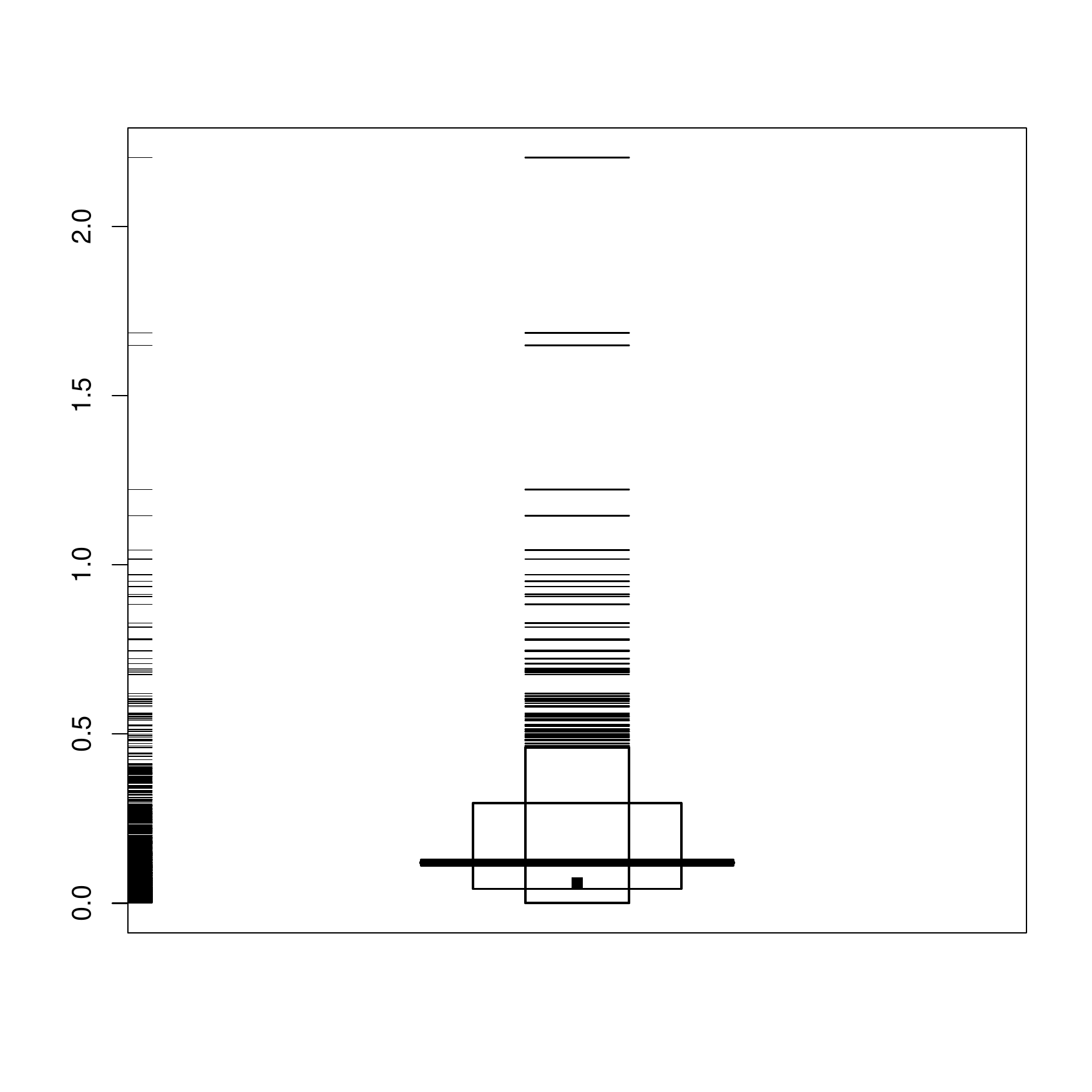}}
\caption{Modified boxplots for the intenssity Niigata VV data set.}\label{Fig3}
\end{center}
\end{figure}

Figure~\ref{Fig3} presents these last three graphical representations of the Niigata VV data set.
Although the visual complexity is somewhat increased from the plots presented in Fig.~\ref{Fig2}, the amount of visual information is also enhanced.

\citeasnoun{Beanplot} introduced the beanplot,  an alternative to the boxplot for visual comparison of univariate data between groups. 
In this plot,  individual observations are shown as small lines in a one-dimensional scatter plot, and then a density estimator of data distribution and the mean are showed. 
An interesting feature of this plot is that bimodalities and possible duplicated measurements are easily exposed. 
{\tt R} has the \verb|beanplot| package which includes a function for this purpose.

Finally,  \citeasnoun{ViolinPlot} presented the violin plot: a combination of a boxplot and a (doubled) kernel density plot. 
The violin plot does not include the individual points, but it displays the median and a box indicating the interquartile range. 
It is useful when comparing multiple groups and with large dataset. 
Overlaid on this boxplot is a kernel density estimation plot. 
There is also a {\tt vioplot} package in \texttt R. 

\begin{figure}[hbt]
\begin{center}
\subfigure[Bean plot ]{\includegraphics[ width=.32\linewidth]{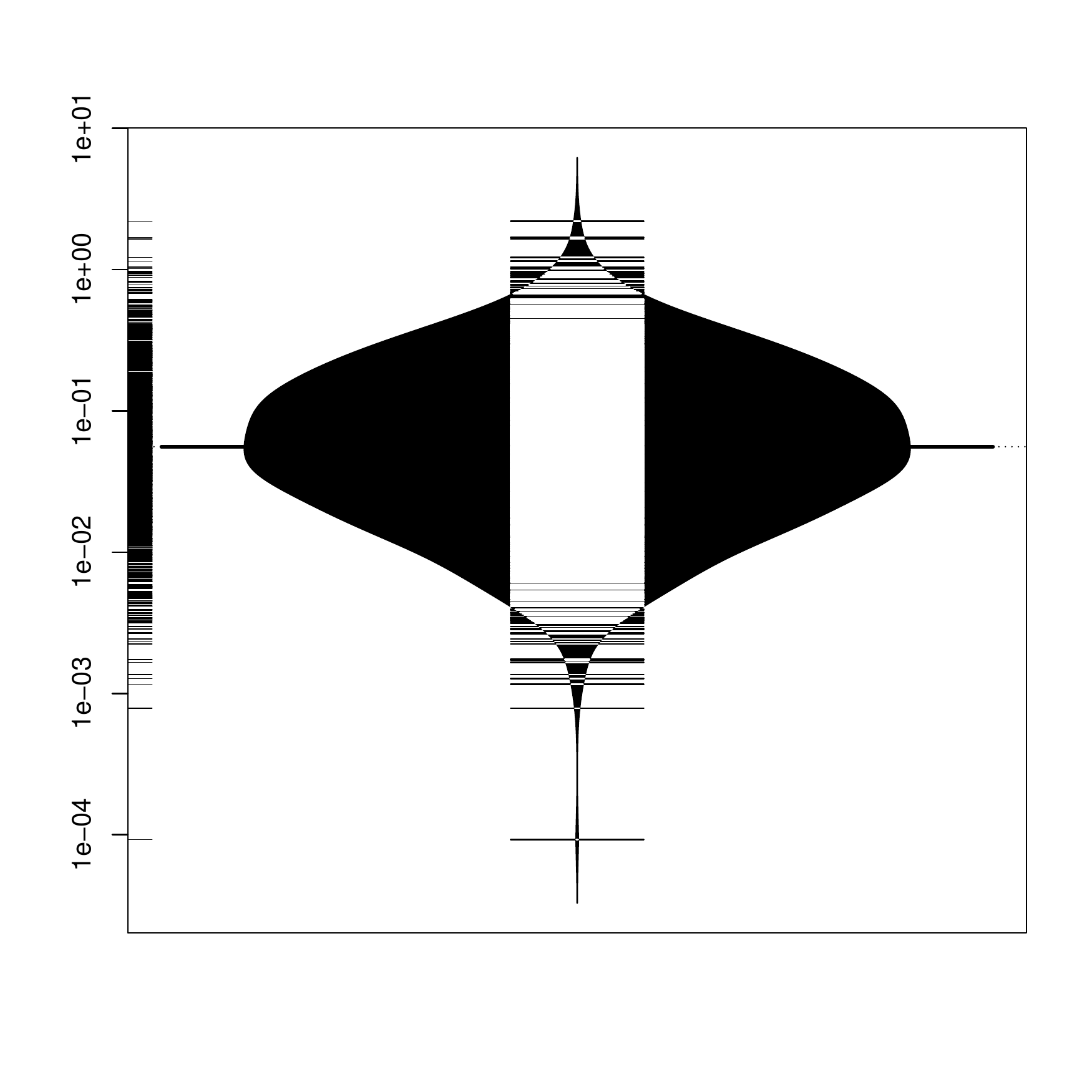}}
\subfigure[Violin plot]{\includegraphics[width=.32\linewidth]{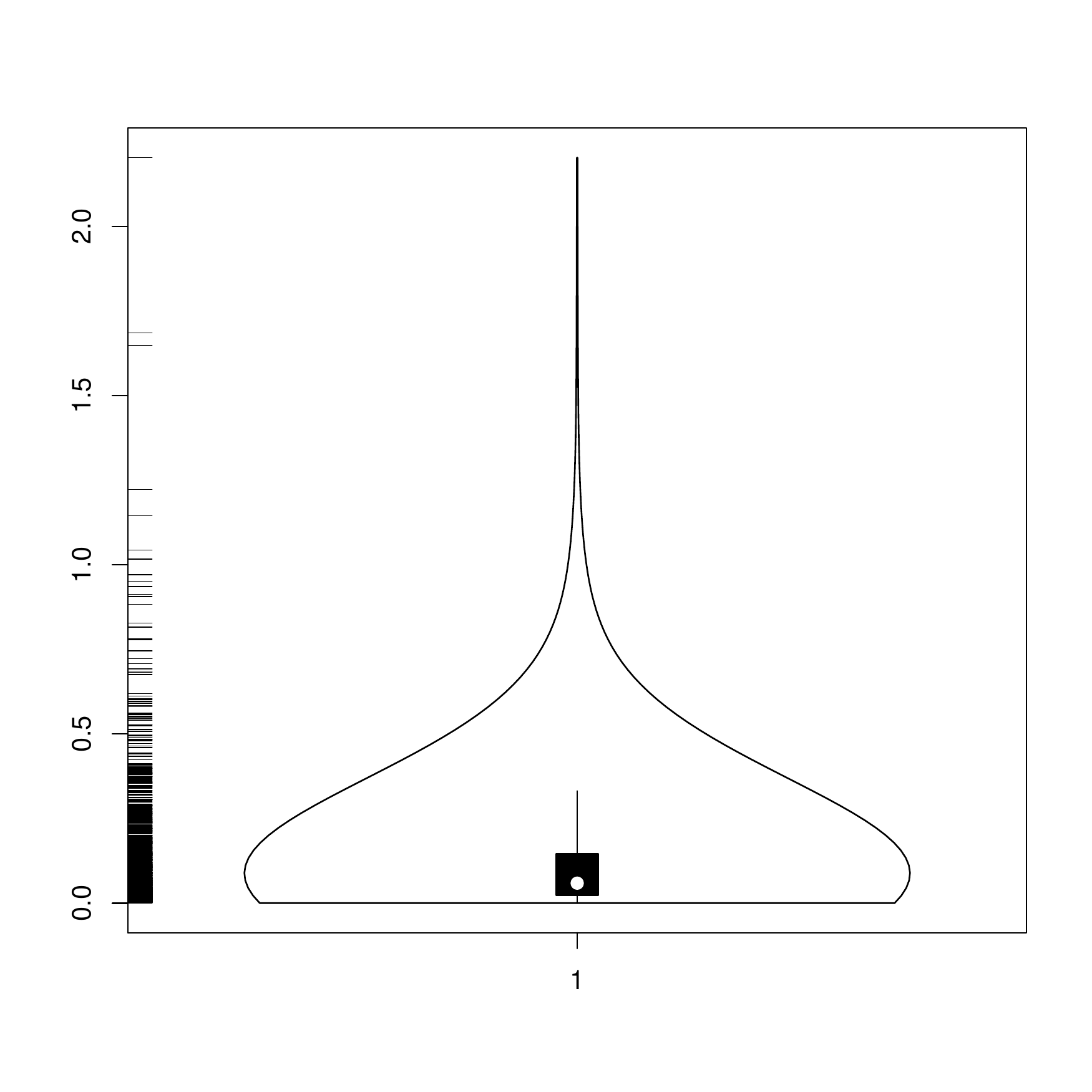}}
\caption{Bean plot and violin plot for the intenssity VV Niigata data set.}\label{Fig4}
\end{center}
\end{figure}

Fig.~\ref{Fig4} presents the bean plot and the violin plot for the Niigata VV data set.

\subsection{The proposed tool}

We discussed in the previous section a number of graphical summaries.
Each conveys an important aspect of the properties of the sample.
Our proposal consists of the simultaneous use of them in a synchronized fashion, for an enhanced visualization of those different aspects.

In order to assess more adequately data whit highly asymmetry and to extract the largest possible amount of information and features from the data data set, we propose the simultaneous use of the histogram enhanced with density estimation, the boxplot with notches (B-N) , the violin plot (V-P), the shifting boxplot (S-P), the adjusted boxplot (A-B), and the box-percentile plot (B-P). 
All graphical summaries are synchronized with respect to a location parameter estimate.
The proposed tool follows the guidelines proposed by \citeasnoun{Tufte01} for quality visual display of quantitative information, in particular, the ratio information/ink was maximized by showing only graphical elements which convey essential information.

The median is shown by as a vertical line connecting all plots 
A rug plot is displayed in the middle of the graphical summaries to reinforce the position of the data, and to reinforce the visual notion that the data is common to all plots. 

\begin{figure}[htb]
  \centering
  \includegraphics[angle=-90,width=\linewidth]{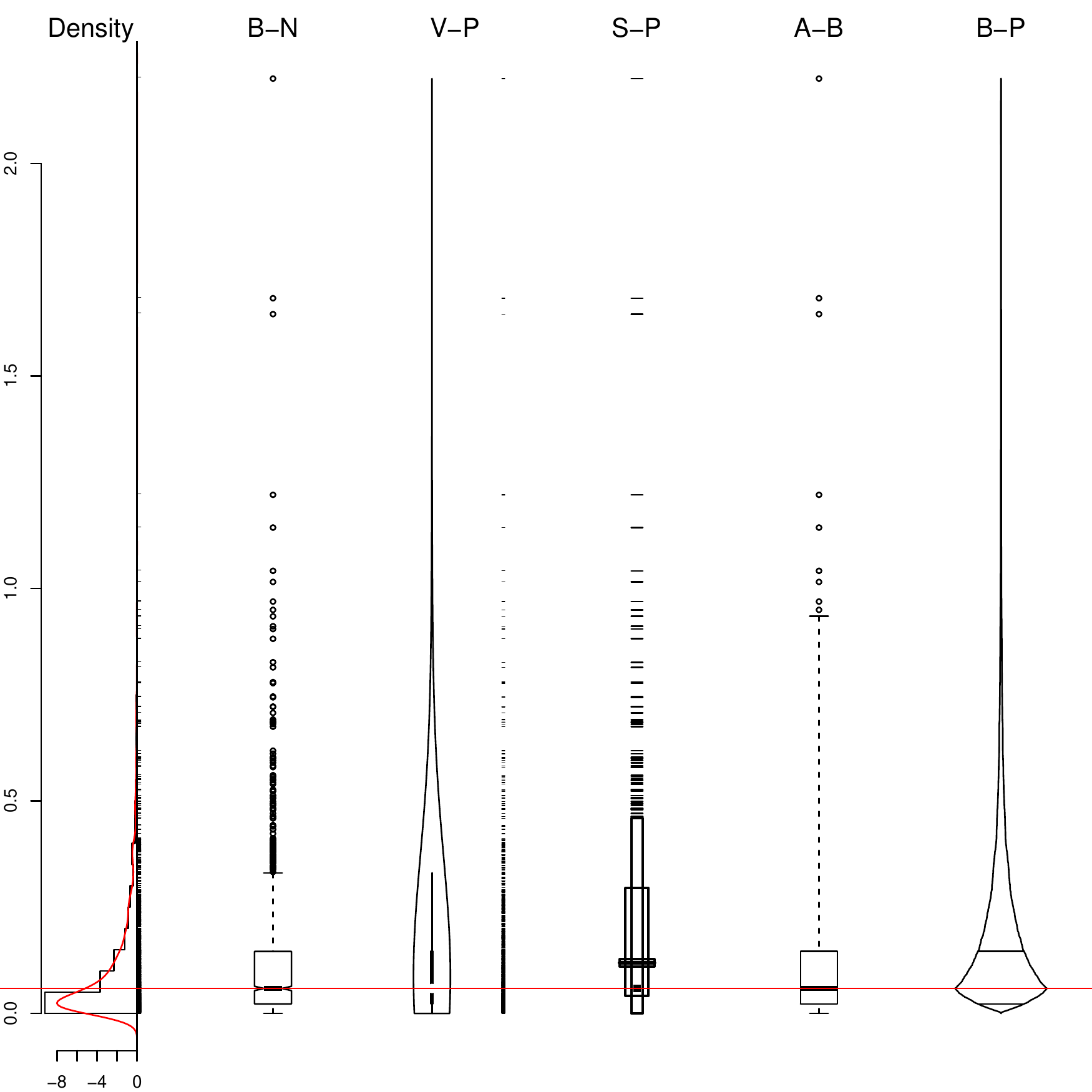} 
  \vspace{0.5cm}\caption{Synchronized graphs in the median of intensity VV of Niigata data set.}\label{Fig5}
\end{figure}

Fig.~\ref{Fig5} presents the result of computing the proposed visualization on the Niigata VV data set.

\section{Application to SAR data}\label{Sec:Results}

SAR sensors provide relevant and complementary information about the target.
Their use has proven valuable from as diverse applications as the mapping of the surface of Venus by the Magellan and Venera missions~\cite{MagellanVenera}, the unearthing of lost Maya ruins~\cite{RadarMappingArcheologyAncientMayaLandUse}, and the 4-D (space and time) monitoring of the environment~\cite{TutorialSAR}.

The main characteristics that make valuable the data provided by SAR sensors are (i)~their ability to provide images with high spatial resolution independent from daylight, cloud coverage and weather and environmental conditions as fog, smoke, smog, rain etc., and (ii)~the fact that their return is the result of complex interactions between the incident signal and the target, which complement the information available in the visible and near-visible spectrum.
The recent tutorial by \citeasnoun{TutorialSAR} is an excellent starting point for the reader interested in this field.

The simplest form SAR data adopt is the intensity.
\citeasnoun{MejailJacoboFreryBustos:IJRS} presented evidence that the $\mathcal G_I^0$ model is able to describe many types of target textures, from textureless (such as crops) to extremely textured (as, for instance, urban areas), but including areas with moderate texture (e.g. forests).
This model had been proposed by \citeasnoun{frery96}, and was later extended to the full polarimetric case by \citeasnoun{FreitasFreryCorreia:Environmetrics:03}.
The $\mathcal G_I^0$ distribution has positive support, and it is indexed by three parameters: the number of looks, which describes the signal-to-noise ratio, the scale, which amounts for the energy incident in the receiving antenna relative to the emitted power, and the texture.
Depending on a relationship between the two last parameters, the $r$-th order moment of this this distribution is infinite.

The visualization of both the density and data from this distribution can be demanding, since extreme observations  are expected.
Among the works which present qualitative analyses of SAR data, \citeasnoun{FreryCorreiaFreitas:ClassifMultifrequency:IEEE:2007} and \citeasnoun{AutomatedNonGaussianClusteringPolSAR} make critical decisions based on such information.
The former decides which joint distribution will be used for the classification, while the latter forms a stopping rule for the iterative segmentation.

\begin{figure}[!hbt]
\centering
\subfigure[San Francisco, sample in yellow\label{San_Francisc}]{\includegraphics[ width=.48\linewidth]{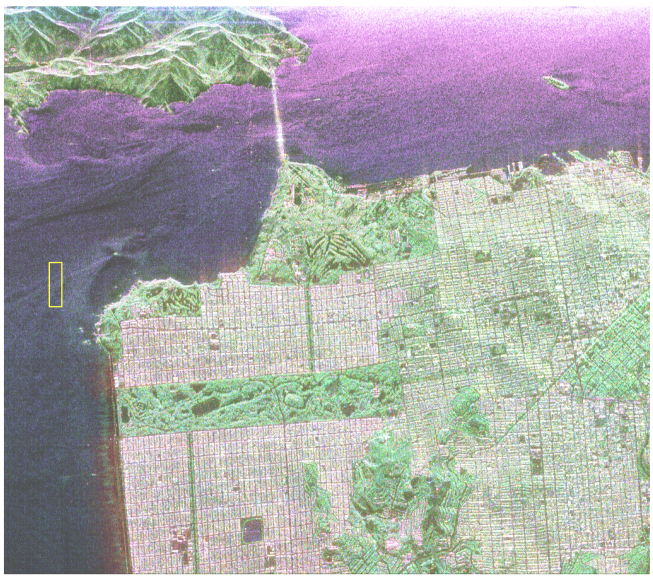}}
\subfigure[Death Valley, sample in red\label{Death}]{\includegraphics[viewport=1 1 367 320 , clip=TRUE, width=.48\linewidth]{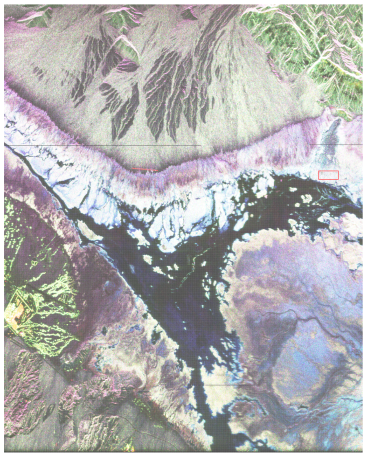}}
\subfigure[Niigata, sample in yellow\label{Niigata}]{\includegraphics[width=.48\linewidth]{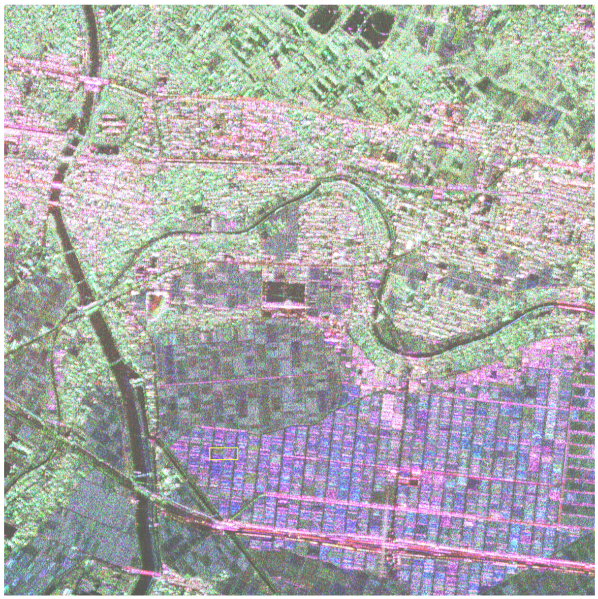}}
\caption{Color composites and samples}\label{fig:images}
\end{figure}

In the following we present the data that will be analyzed with the proposed technique.
Fig.~\ref{fig:images} presents the color composites of three images from different PolSAR sensors and areas.
Fig.~\ref{San_Francisc} is from the San Francisco area, and the area under analysis is highlighted in yellow; it is a textureless sample from the sea.
Fig.~\ref{Death} is from the Death Valley, and the sample in red has moderate texture.
Fig.~\ref{Niigata} is from Niigata, and the sample in yellow has extreme texture since it is from an urban area.

Table~\ref{tab:Summary} presents quantitative summary information about these data sets.
As expected, regardless the range, the image and the polarization of the data, there is intense skewness and kurtosis in all cases.

\begin{sidewaystable}[hbt]
\centering\small \caption{Summary statistics.}\label{tab:Summary}
\renewcommand{\arraystretch}{1.5}
\begin{tabular}
           {           
           @{}
           lp{1.2cm}p{1.2cm}p{1.2cm}lllp{1.2cm}p{1.2cm}p{1.2cm}
           S[table-format=3.2]
           S[table-format=3.2]
           S[table-format=3.2]
           S[table-format=3.2]
           S[table-format=3.2]
           S[table-format=3.2]
           S[table-format=3.2]
           S[table-format=3.2]
           S[table-format=3.2]
           S[table-format=3.2]
           S[table-format=3.2]
           S[table-format=3.2]
           @{}
           }
\toprule
& \multicolumn{3}{c}{San Francisc} & \multicolumn{3}{c}{Death Valley} &  \multicolumn{3}{c}{Niigata}\\
& \multicolumn{3}{c}{$N=4320$} & \multicolumn{3}{c}{$N=4446$} &  \multicolumn{3}{c}{$N=4446$}\\
\cmidrule(r){2-4}
\cmidrule(r){5-7}
\cmidrule(r){8-10}
         & HH & HV & VV & HH & HV & VV & HH & HV & VV \\
\cmidrule(r){2-4}
\cmidrule(r){5-7}
\cmidrule(r){8-10}
Min      & 1.45$\times 10^{-4}$ & 1.04$\times 10^{-4}$ & 8.92$\times 10^{-4}$ & 0.72 & 0.03 & 0.27 & 1.15$\times 10^{-4}$ & 5.37$\times 10^{-7}$ & 9.22$\times 10^{-5}$\\
1st Quartile & 3.48$\times 10^{-3}$ & 5.04$\times 10^{-4}$ & 0.01 & 0.44 & 0.13 & 0.92 & 0.02 & 1.97$\times 10^{-3}$ & 0.02\\
Median   & 5.53$\times 10^{-3}$ & 7.64$\times 10^{-4}$ & 0.02 & 0.56 & 0.18 & 1.18 & 0.05 & 4.54$\times 10^{-3}$  & 0.06\\
Mean     & 6.64$\times 10^{-3}$ & 8.91$\times 10^{-4}$ & 0.02 & 0.58 & 0.18 & 1.23 & 0.10 & 0.01 & 0.12\\
3rd Quartile & 8.62$\times 10^{-3}$ & 1.13$\times 10^{-3}$ & 0.03 & 0.69 & 0.22 & 1.49 & 0.11 & 0.01 & 0.15\\
Max      & 0.04 & 3.83$\times 10^{-3}$ & 0.13 & 1.45 & 0.48 & 3.15 & 1.42 & 0.04 & 2.20\\
Skeweness   & 1.84 & 1.56 & 1.57 & 0.71 & 0.66 & 0.62 & 3.74 & 1.92 & 4.13\\
Kurtosis & 9.34 & 6.60 & 7.49 & 3.59 & 3.77 & 3.64 & 23.37 & 7.78 & 32.12\\
\bottomrule
\end{tabular}
\end{sidewaystable}

\begin{sidewaysfigure}[!hbt]
\centering
\subfigure[Intensity HH\label{ihhSF}]{\includegraphics[angle=-90, width=.32\linewidth]{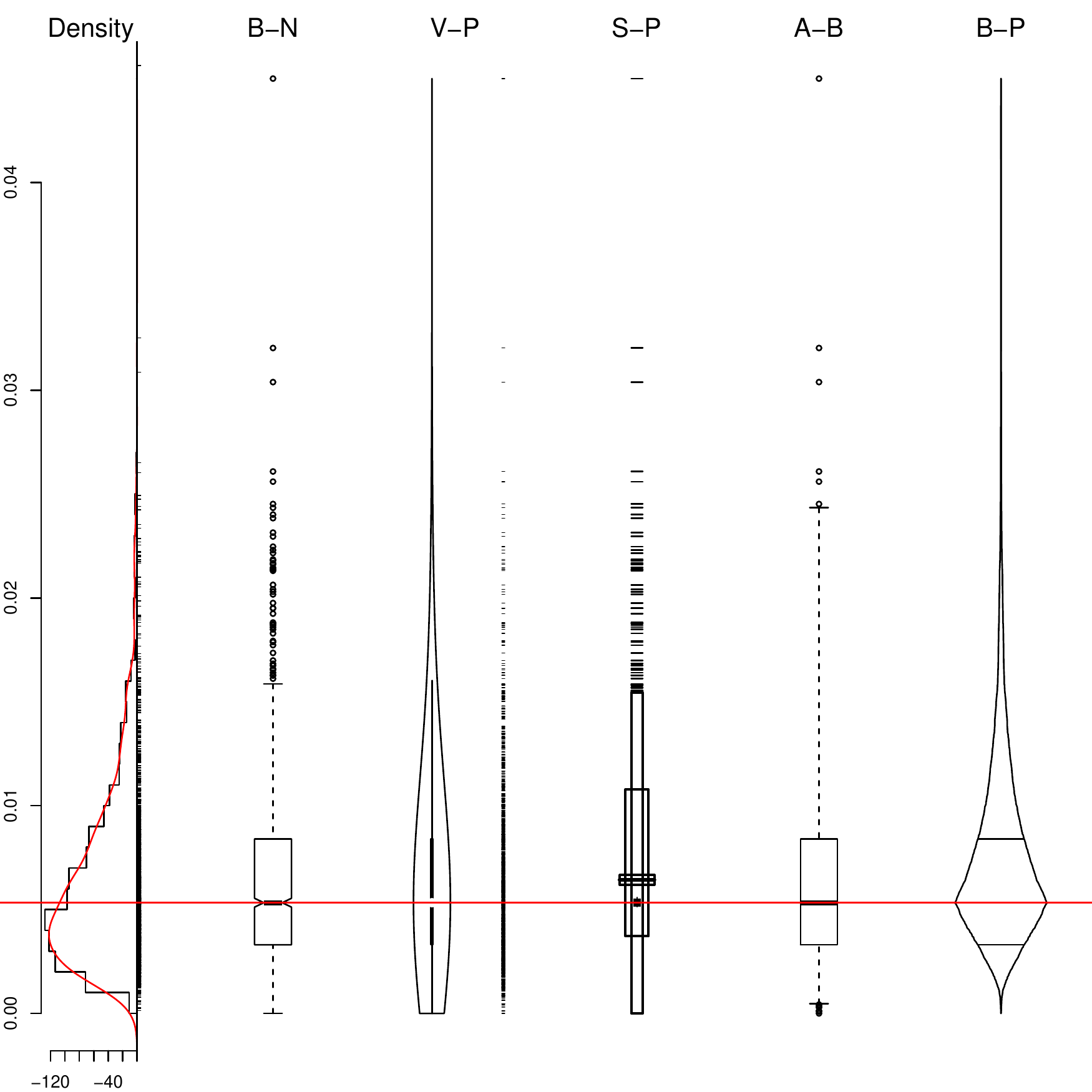}}
\subfigure[Intensity HV\label{ihvSF}]{\includegraphics[angle=-90, width=.32\linewidth]{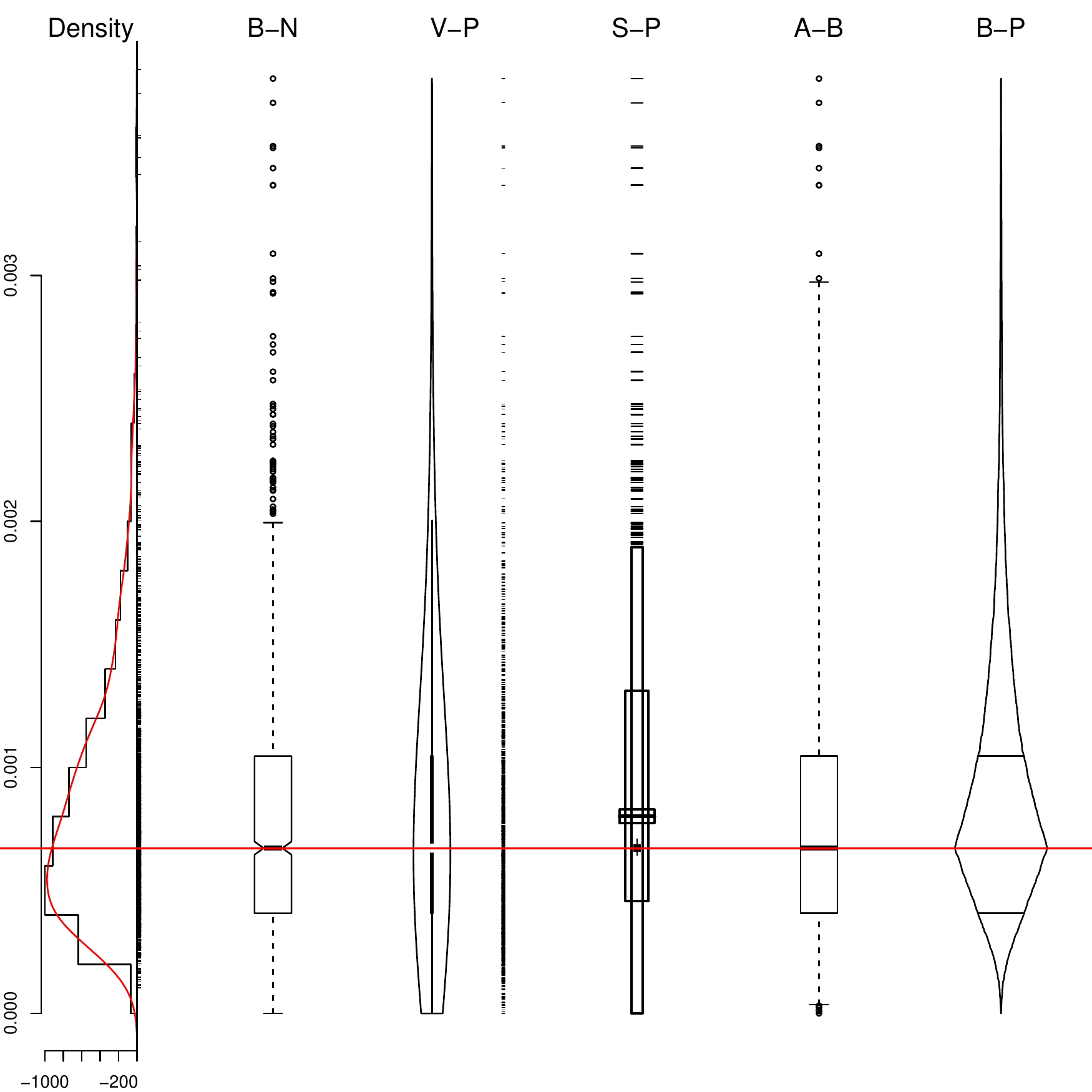}}
\subfigure[Intensity VV\label{ivvSF}]{\includegraphics[angle=-90, width=.32\linewidth]{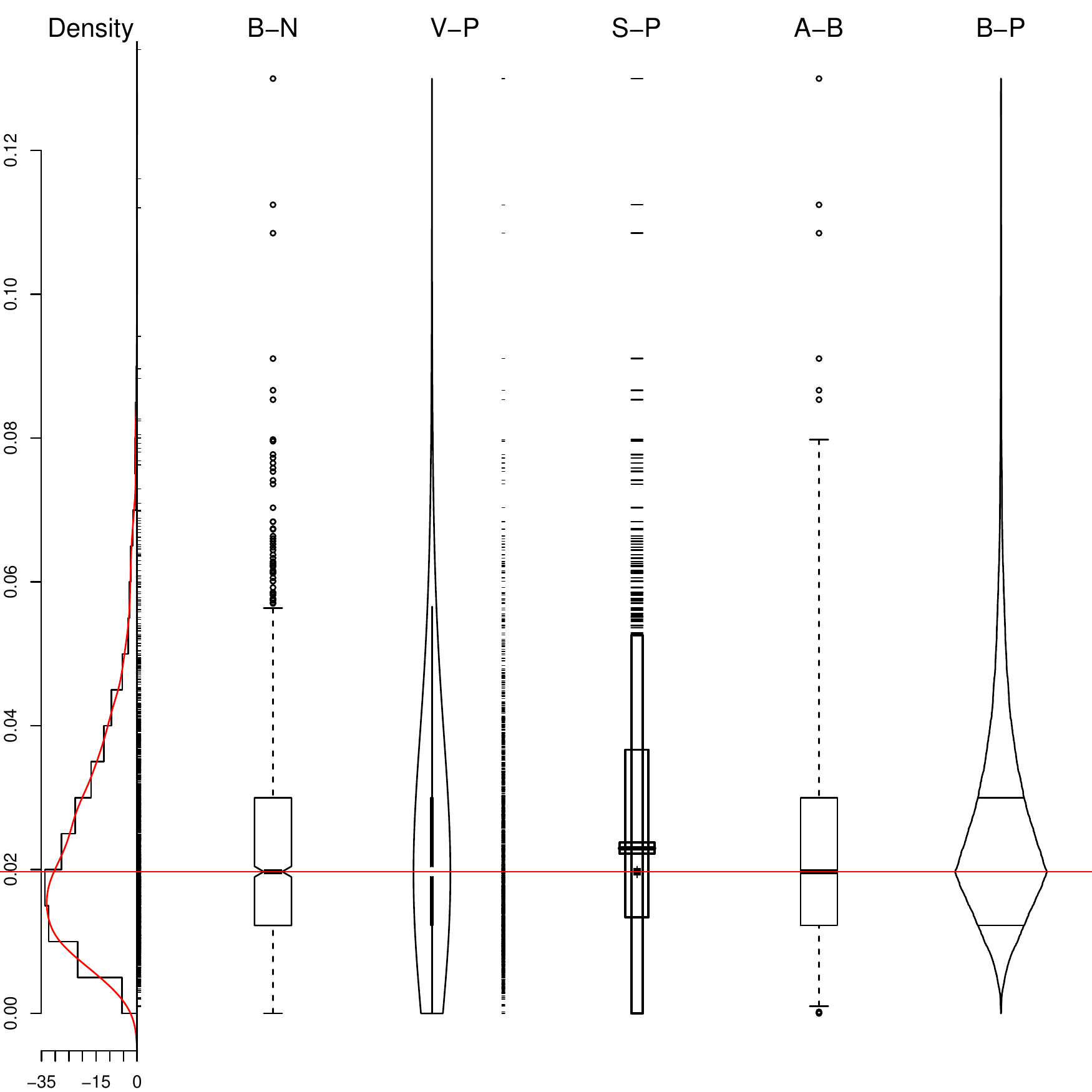}}
\caption{Synchronized graphs in the median of intensity bands of  San Francisco data set.}\label{Fig:SF}
\end{sidewaysfigure}

\begin{sidewaysfigure}[!hbt]
\centering
\subfigure[Intensity HH\label{ihhDV}]{\includegraphics[angle=-90, width=.32\linewidth]{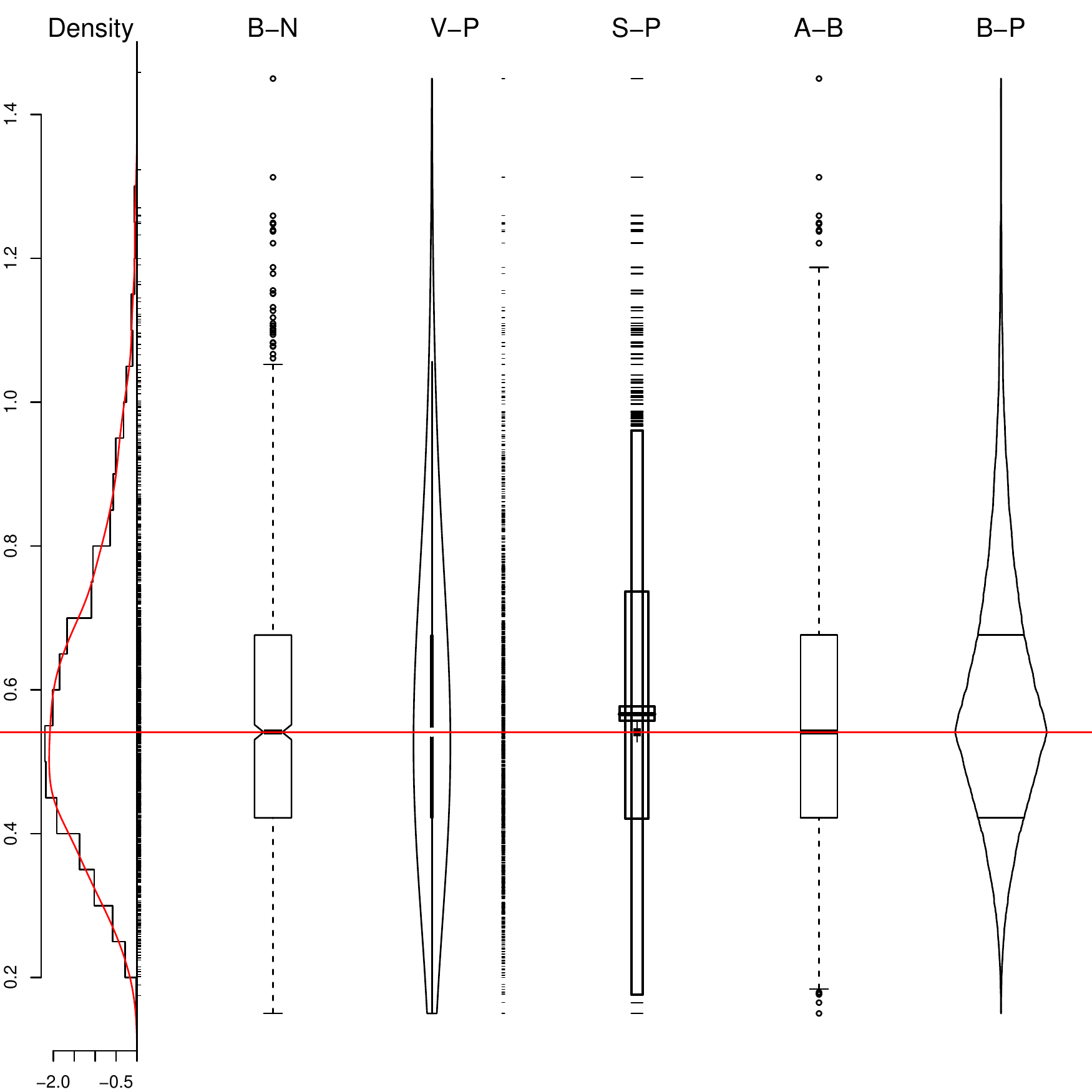}}
\subfigure[Intensity HV\label{ihvDV}]{\includegraphics[angle=-90, width=.32\linewidth]{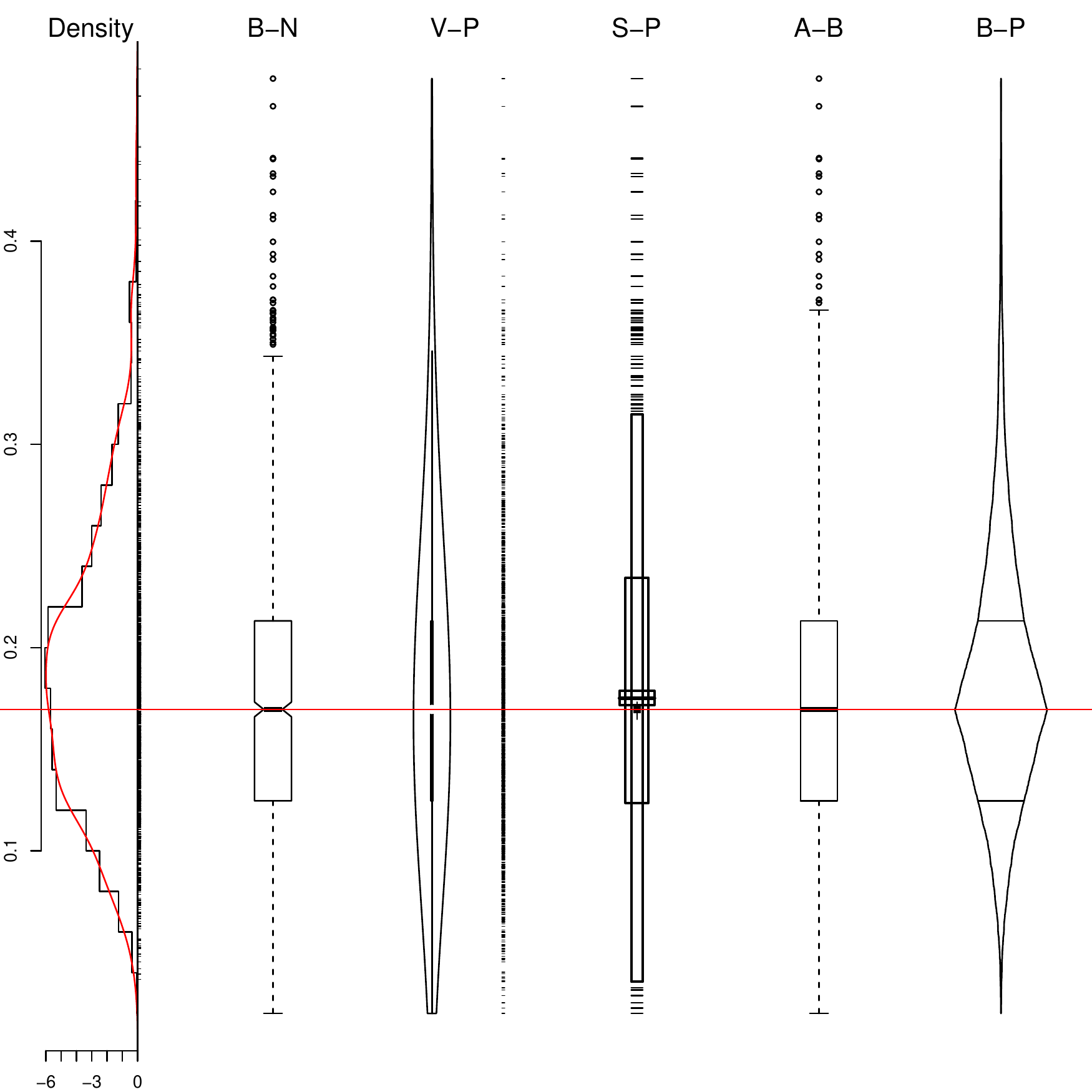}}
\subfigure[Intensity VV\label{ivvDV}]{\includegraphics[angle=-90, width=.32\linewidth]{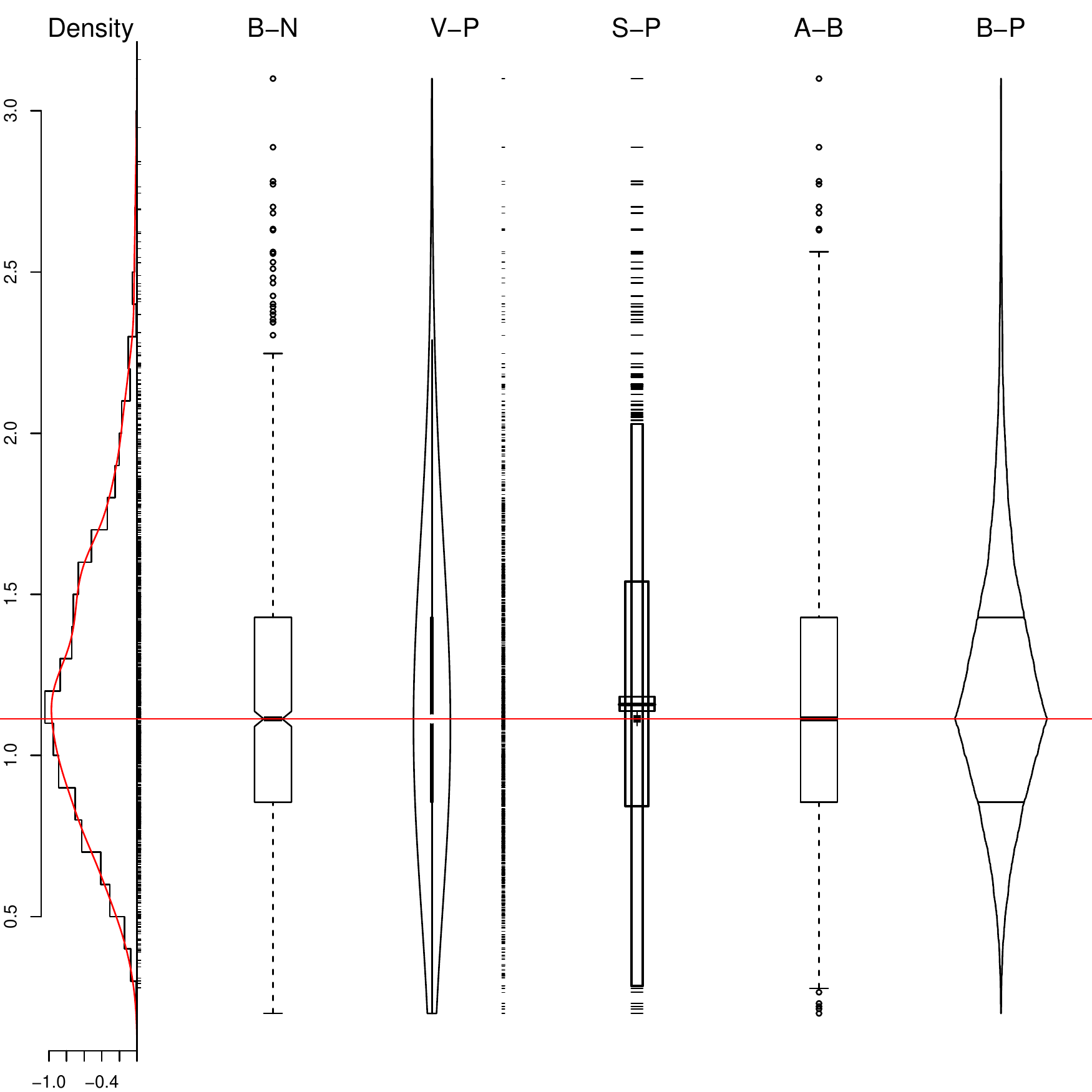}}
\caption{Synchronized graphs in the median of intensity bands of Death Valley data set.}\label{Fig:DV}
\end{sidewaysfigure}

\begin{sidewaysfigure}[!hbt]
\centering
\subfigure[Intensity HH\label{ihhNGtemp}]{\includegraphics[angle=-90, width=.32 \linewidth]{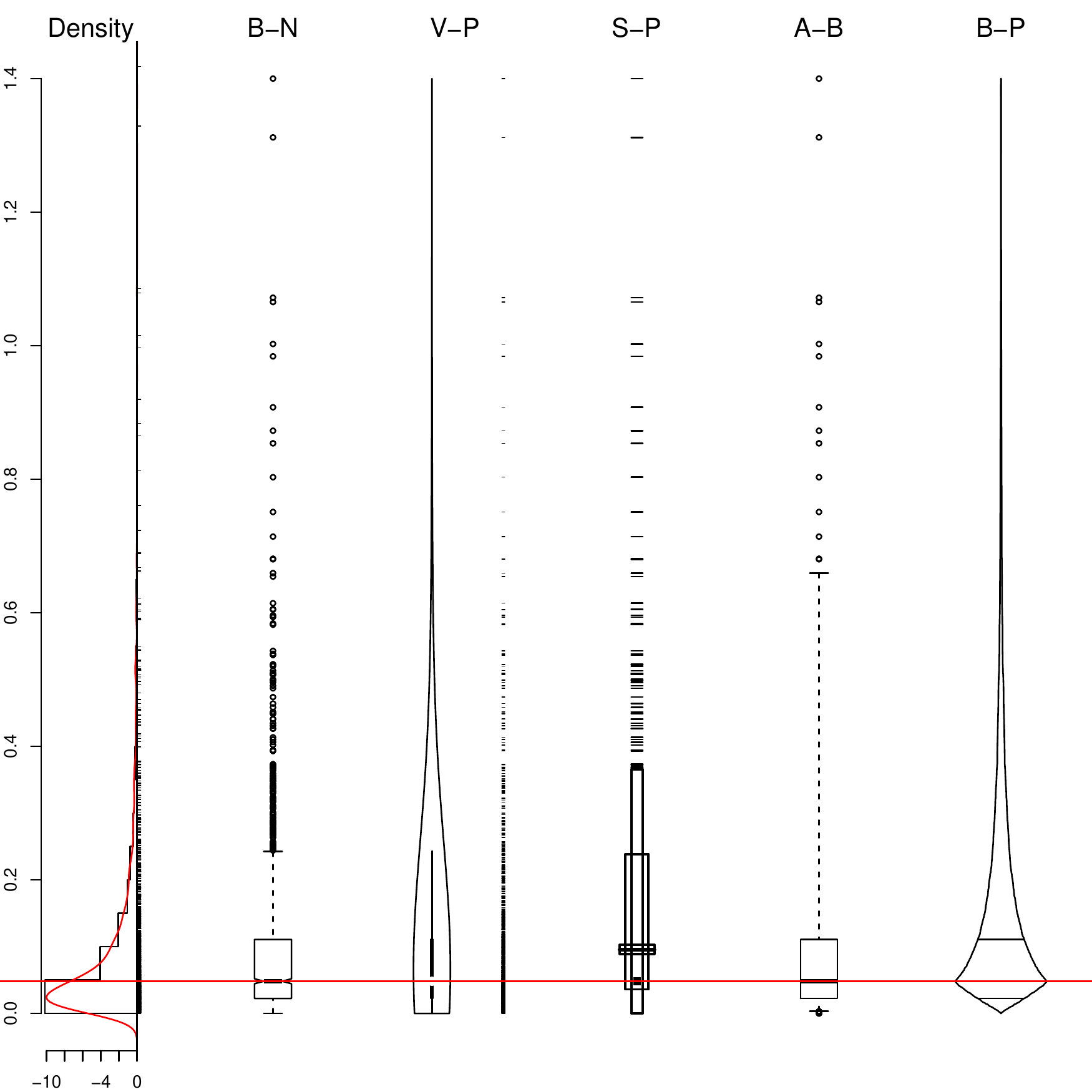}}
\subfigure[Intensity HV\label{ihvNGtemp}]{\includegraphics[angle=-90, width=.32\linewidth]{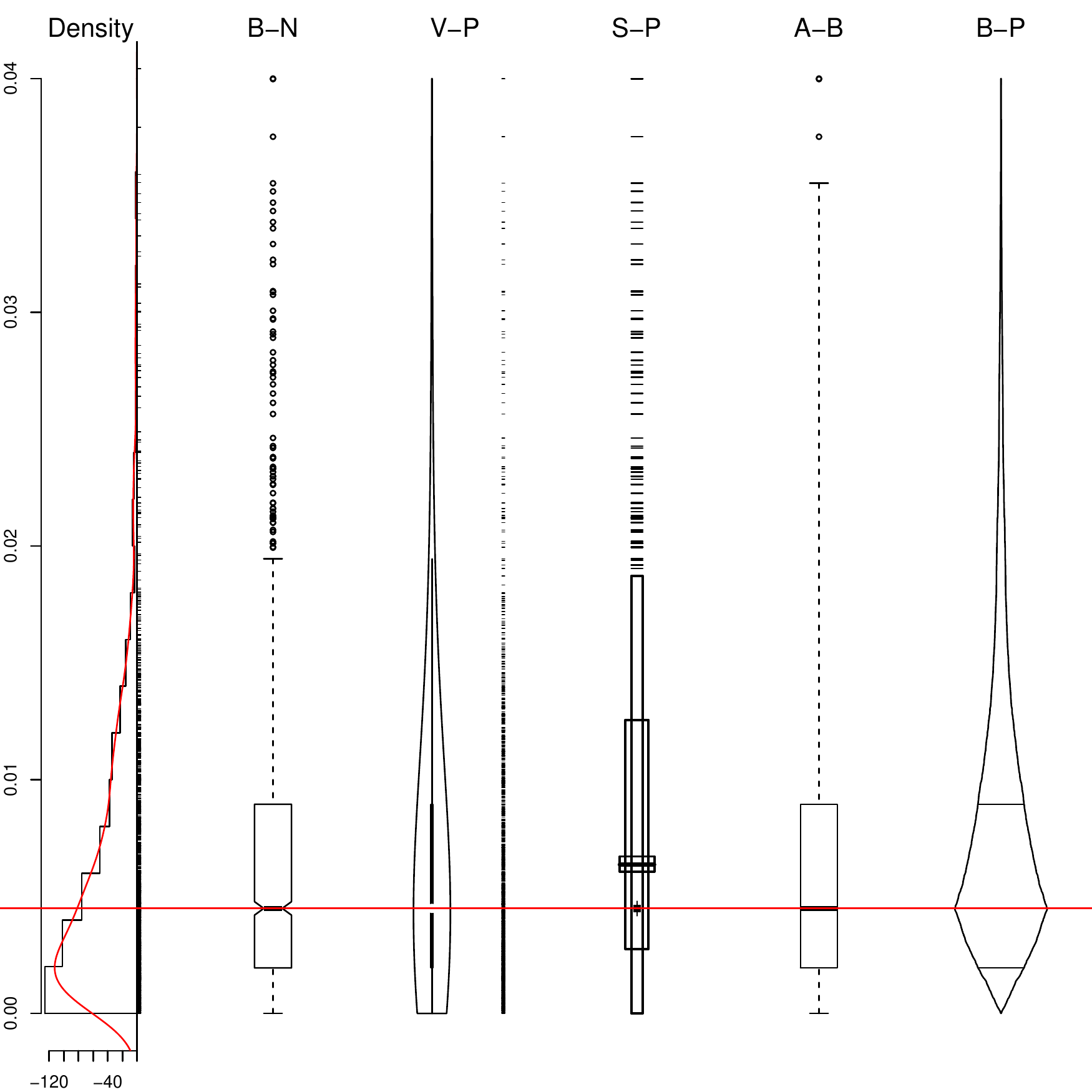}}
\subfigure[Intensity VV\label{ivvNGtemp}]{\includegraphics[angle=-90, width=.32\linewidth]{Synchronized-ivvNG.pdf}}
\caption{Synchronized graphs in the median of intensity bands of Niigata data set.}\label{Fig:NGtemp}
\end{sidewaysfigure}

Figures~\ref{Fig:SF}, \ref{Fig:DV} and~\ref{Fig:NGtemp} show the results of applying the proposed technique to each polarization of the samples from figures~\ref{San_Francisc}, \ref{Death} and~\ref{Niigata}, respectively.
As can be seen from the values presented in Table~\ref{tab:Summary}, these data sets are comparable within each image, but the ranges differ widely among images.


The samples presented in Fig.~\ref{Fig:SF} look alike when comparing the histograms and fitted densities, but important differences arise in the violin and shifting boxplots.
In these graphical summaries is clearer that the HV band has more spread than the other two.
The extent can be visually quantified by the shifting boxplots.


The data from the Death Valley image, summarized in Fig.~\ref{Fig:DV}, are the most symmetric; this confirms the values presented in Table~\ref{tab:Summary}.
Nevertheless, one observes in the boxplots that there are outliers to the right of the three samples.
If asymmetry is assumed, the adjusted boxplot also detects ouliers to the left of two of the three polarizations, namely in figures~\ref{ihhDV} and~\ref{ivvDV}.


Fig.~\ref{Fig:NGtemp} shows at a glance the different behavior among polarizations, with the HV channel exhibiting more spread than the other two; cf.\ the box-percentile and violin plots
Although HH and VV polarizations behave alike, the adjusted boxplot reveals that the former has more outliers than the latter.

\section{Conclusions and future work}\label{Sec:Conclu}

As presented in the examples, the use of synchronized graphical summaries promoted the discovery of information conveyed by the data.
Had only one type of plot been used, some of these features would have not been identified.

Synchronization is essential for retaining the ability to compare graphical representations of the same data set.
A loose presentation of two or more of the plots would not allow the discovery of such information.

More customizable options are being added to the tool as, for instance, the ability to choose interactively the order in which the plots appear.

In its current version, the function does not return any object.
This can be easily customized, for instance using lists.


\section*{ Acknowledgments}

The authors wish to thank Dr.\ Fernando Marmolejo for providing the {\tt R} codes of the shifting boxplot.  
The study was supported partially by CNPq and Fapeal grants, from Brazil.

\bibliography{references}
\appendix
\section{Implementation}

The tool for producing synchronized plots was written in the \texttt R programming language~\cite{RSoftware}.
The code involves functions from other freely available packages: {\tt Hmisc, robustbase, vioplot, bootstrap, MASS, lfstat, graphics, gplots, beanplot}, and some new implementations for horizontal histogram, lines density estimates in violin plots, and  Box-plot percentile.

The proposed new {\tt R} function, termed {\tt SincronizedPlot}, joins in synchronized form the histogram enhanced with a density estimation, the boxplot with notches (B-N), the violin plot (V-P), the shifting boxplot (S-P), the adjusted boxplot (A-B), and the box-percentile plot (B-P). 
The main argument that must be supplied is
{\tt data}, a numeric vector or an {\tt R} object which is coercible to one by \verb|'as.vector(x, "numeric")'|

The time required to produce an output is negligible.

The code and data used in this work are freely available from \url{http://www.de.ufpe.br/~raydonal/SynchronizedPlots/SynchronizedPlots.zip}.
In order to try the tool, the user must load the {\tt R} scripts, and issue the following {\tt R} commands:
\begin{verbatim}
# Data
x=rgamma(100, shape=0.5)
# Plot
SincronizedPlot(x)
\end{verbatim}

\end{document}